\documentclass[prd,aps,twocolumn,nofootinbib]{revtex4}

\usepackage{graphicx}

\begin{document}

\title{Axion isocurvature fluctuations with extremely blue spectrum}

\author{Shinta Kasuya$^a$ and Masahiro Kawasaki$^{b,c}$}

\affiliation{
$^a$ Department of Information Science,
     Kanagawa University, Kanagawa 259-1293, Japan\\
$^b$ Institute for Cosmic Ray Research,
     University of Tokyo, Chiba 277-8582, Japan\\
$^c$ Institute for the Physics and Mathematics of the Universe, 
     University of Tokyo, Chiba 277-8582, Japan}

\date{April 24, 2009}

\begin{abstract}
We construct an axion model for generating isocurvature fluctuations with blue spectrum, 
$n_{\rm iso}=2-4$, which is suggested by recent analyses of admixture of adiabatic and 
isocurvature perturbations with independent spectral indices, $n_{\rm ad} \ne n_{\rm iso}$. 
The distinctive feature of the model is that the spectrum is blue at large scales while 
scale invariant at small scales. This is naturally realized by the dynamics of the 
Peccei-Quinn scalar field.
\end{abstract}


\maketitle


\section{Introduction}
Large-scale structures of the universe, such as galaxies and clusters of galaxies, have 
formed through gravitational instabilities, initiated by the primordial seed density fluctuations,
which were created during inflation. The simplest initial condition seeded these
inhomogeneities is the (almost) scale-invariant adiabatic curvature perturbations. They can
fit to very precise measurements of the cosmic microwave background 
temperature and polarization anisotropies, large-scale structures, 
and supernovae \cite{WMAP,SDSS,SN}. It is usually realized by the single-field inflation 
where the inflaton fluctuations are responsible for the adiabatic perturbations.

Generally, there will exist other light fields whose fluctuations during inflation become
isocurvature perturbations \cite{iso}. Therefore, the admixture of isocurvature and adiabatic 
fluctuations could be what really happened in the early universe. Observational analyses
with the assumption that the spectral indices of the adiabatic and isocurvature fluctuations are
the same, $n_{\rm ad} = n_{\rm iso}\simeq 1$, have revealed that the contributions from the 
isocurvature perturbation should be small \cite{WMAP,iso_analysis}.

However, there is {\it a priori} no reason for the isocurvature fluctuations to have (almost) 
scale-invariant spectrum. In fact, more general analyses with independent spectral indices
of adiabatic and isocurvature modes based on recent observations result in the favor of 
much more contribution of the isocurvature component with an extremely blue 
tilt ($n_{\rm iso}\simeq 2-4$) \cite{blue1,blue2,blue3}.

In this article, we provide a concrete model for generating isocurvature fluctuations with an 
extremely blue spectrum for the first time.\footnote{
The possibility to obtain isocurvature fluctuations with some deviation from scale-invariant was
investigated in Ref.~\cite{Linde}.}
It is the axion model in supersymmetry (SUSY) \cite{KKY}.
Since the axion is a good candidate of the cold dark matter of the universe, and has nothing
to do with radiation, it gives rise to uncorrelated isocurvature fluctuations. The axion isocurvature
fluctuation is usually expressed as $\delta a /a \simeq H/(2\pi F_a \theta)$, where $H$ is the 
Hubble parameter during inflation, $F_a$ the axion decay constant, and $\theta$ a misalignment 
angle. The key to produce the blue spectrum is that we promote $F_a$ as a dynamical 
field $\varphi\equiv``F_a"$, which initially has a large value $\simeq M_P$, evolves 
toward smaller values, and stops at $\simeq F_a$ during inflation. It is realized very 
simply and naturally in the SUSY axion model, and we can obtain extremely blue 
spectrum such as $n_{\rm iso} \simeq 4$ at large scales, which is connected 
to the scale-invariant spectrum at small scales. 

The structure of the article is as follows: In the next section, we explain the essence to generate
the extremely blue spectrum in a simple model, reduced from the concrete model that 
we provide based on SUSY in Sec.III. We then show the dynamics of the fields which leads 
to the favorable spectrum in Sec.IV. Our conclusions are given in Sec.V.

\section{How to get the blue spectrum}
Let us consider a toy model of a complex scalar field $\Phi$, whose energy density is negligible 
during inflation. Fluctuations in the phase 
direction give rise to an isocurvature perturbation, while fluctuations in the radial direction 
are negligibly small due to large effective mass in that direction as shown shortly. 
Thus, the isocurvarture fluctuation is given by
\begin{equation}
\frac{\delta \theta}{\theta}  \simeq \frac{H}{2\pi \varphi \theta},
\end{equation}
where we denote $\Phi=\varphi e^{i\theta}/\sqrt{2}$. Since the Hubble parameter during 
inflation is (almost) constant, it is the decreasing amplitude of $\varphi$ that makes the 
isocurvature perturbation blue tilted. When the field $\varphi$ has mass of $O(H)$, it can 
roll down in the potential during inflation, and, in addition, its fluctuation $\delta\varphi$ 
is suppressed. The reduced potential is given by
\begin{equation}
V  \simeq  \frac{1}{2} c H^2 \varphi^2,
\end{equation}
when $\varphi$ has a large field value, and $c\sim O(1)$ is a constant. 
Then the $\varphi$ field obeys the equation
\begin{equation}
\ddot{\varphi} + 3H\dot{\varphi} + c H^2 \varphi = 0,
\end{equation}
which has a solution of the form $\varphi \propto e^{-\lambda H t}$ with
\begin{equation}
\lambda = \frac{3}{2} - \frac{3}{2}\sqrt{1-\frac{4}{9}c},
\end{equation}
for $0\le c\le 9/4$.\footnote{
One obtains the damping oscillating solution for $c>9/4$. Since it does not suit for our purpose,
we only consider for $c\le 9/4$.} 
Since the isocurvature fluctuation is estimated as
\begin{equation}
\Delta_{\rm iso}^2 \propto \left(\frac{\delta a}{\varphi}\right)^2 \sim \left(\frac{H}{\varphi}\right)^2
\propto e^{2\lambda H t},
\end{equation}
its spectral index is given by
\begin{equation}
n_{\rm iso} -1 \equiv \frac{d\ln\Delta_{\rm iso}^2}{d\ln k} = 2\lambda = 3 - 3\sqrt{1-\frac{4}{9}c}.
\end{equation}
Therefore, we obtain the blue spectrum, even extremely blue such as $n_{\rm iso} = 4$ for 
$c=9/4$. As shown explicitly in the following sections, the field $\varphi$ eventually settles
down in the minimum of the potential placed at $\varphi \simeq F_a$. Thereafter the 
isocurvature flucutation becomes scale invariant.

\section{Axion model in SUSY}
The axion \cite{PQ,axion} is a Nambu-Goldstone boson associated with the Peccei-Quinn (PQ) 
symmetry breaking, and is the most natural solution to the strong CP problem in QCD \cite{SCP}. 
The PQ symmetry breaking scale $F_a$ is astrophysically and cosmologically constrained within 
the range of $10^{10} - 10^{12}$ GeV \cite{axion-rev}. The axion can be cold dark matter for the 
higher values. 

Let us consider a concrete model of the axion in SUSY. We take the following 
superpotential\cite{KKY}:
\begin{equation}
W=h(\Phi_+ \Phi_- - F_a^2)\Phi_0.
\end{equation}
Here $\Phi_+$, $\Phi_-$, and $\Phi_0$ are chiral superfields with PQ charges
$+1$, $-1$, and $0$, respectively, and $h$ is a constant of $O(1)$. The scalar potential is obtained as
\begin{equation}
\label{pot_susy}
V_{\rm SUSY} = h^2\left|\Phi_+\Phi_- - F_a^2\right|^2 + h^2(|\Phi_+|^2+|\Phi_-|^2)|\Phi_0|^2,
\end{equation}
where we denote the scalar components with the same symbols as the superfields.
One can easily see the existence of the flat direction which satisfies
\begin{equation}
\label{FD}
\Phi_+\Phi_- = F_a^2, \qquad \Phi_0=0.
\end{equation}
In addition, SUSY breaking effects lift the flat direction by soft mass terms
\begin{equation}
V_{\rm m} = m_+^2 |\Phi_+|^2 + m_-^2|\Phi_-|^2+m_0^2|\Phi_0|^2,
\end{equation}
at low energy scales, where $ m_+$, $m_-$, and $m_0$ are of $O({\rm TeV})$, as well as the 
so-called Hubble-induced mass terms during inflation,
\begin{equation}
\label{pot-H}
V_{\rm H} = c_+H^2 |\Phi_+|^2 + c_-H^2|\Phi_-|^2+c_0H^2|\Phi_0|^2,
\end{equation}
where $c_+$, $c_-$, and $c_0$ are positive constants of $O(1)$, which stem from the supergravity 
effects \cite{DRT,Hmass}.\footnote{
The coefficients of the Hubble-induced mass terms are determined as
$c_i \simeq 3(1-y_i)$ $(i=0,\pm)$ for the non-renormalizable coupling in K\"ahler potential 
$\delta K= y_i |\Phi_i|^2|I|^2/M_P^2$, where $I$ is the inflaton and $y_i$'s are the coupling constants.}
Notice that no Hubble-induced A terms will appear due to PQ symmetry.
We assume $H\ll F_a$ in order not to destroy the flat direction (\ref{FD}).
Taking into account the fact that $\Phi_0 =0$ and $m_i\ll H \, (i=\pm,0)$ 
during inflation, we only consider the potential of the form
\begin{equation}
\label{pot}
V = h^2\left|\Phi_+\Phi_- - F_a^2\right|^2 + c_+H^2 |\Phi_+|^2 + c_-H^2|\Phi_-|^2.
\end{equation}
Owing to the Hubble-induced mass terms, the minimum of the potential is given by 
\begin{equation}
|\Phi_+^{\rm min}| \simeq  \left(\frac{c_-}{c_+}\right)^{1/4}\!\!F_a, \quad
|\Phi_-^{\rm min}| \simeq  \left(\frac{c_+}{c_-}\right)^{1/4}\!\!F_a.
\end{equation}
Since it is symmetric between $\Phi_+$ and $\Phi_-$, we consider $|\Phi_+| > |\Phi_-|$ without loss
of generality.\footnote{
Notice that, as shown shortly, the amplitude of the isocurvature fluctuation is determined by the 
larger between $|\Phi_+|$ and $|\Phi_-|$, so the spectrum cannot be red tilted.}

Now we must identify the axion field $a$. Rewriting $\Phi_\pm$ as
\begin{equation}
\label{field}
\Phi_\pm \equiv \frac{1}{\sqrt{2}}\varphi_\pm \exp\left(i \theta_\pm \right) , \quad
\theta_\pm \equiv \frac{a_\pm}{\sqrt{2}\varphi_\pm} , 
\end{equation}
we can define the fields $a$ and $b$ as
\begin{eqnarray}
a & = & \frac{\varphi_+}{(\varphi_+^2+\varphi_-^2)^{1/2}}a_+ 
- \frac{\varphi_-}{(\varphi_+^2+\varphi_-^2)^{1/2}}a_-, \\
b & = & \frac{\varphi_-}{(\varphi_+^2+\varphi_-^2)^{1/2}}a_+ 
+ \frac{\varphi_+}{(\varphi_+^2+\varphi_-^2)^{1/2}}a_-.
\end{eqnarray}
From Eqs.(\ref{pot_susy}) or (\ref{pot}), the potential $V(b)$ for the field $b$ is obtained as
\begin{equation}
V(b) = - h^2 F_a^2 \varphi_+ \varphi_- 
\cos\left( \frac{(\varphi_+^2+\varphi_-^2)^{1/2}}{\varphi_+\varphi_-} b \right),
\end{equation}
which implies that the mass of $b$ is given by $\sim h(\varphi_+^2+\varphi_-^2)^{1/2}$. Since 
the field value is $\varphi_+ \simeq M_P$ initially, and decreases until it reaches to $F_a$, as 
shown in the next section, $m_b \gg H$ during inflation and hence the $b$ field quickly settles 
down into the minimum of the potential. On the other hand, the potential for the $a$ field is flat 
and we can regard it as the axion. During inflation, the quantum fluctuations of $a$ develop as
\begin{equation}
\label{delta_a}
\delta a \simeq \delta a_+ \simeq \frac{H}{2\pi},
\end{equation}
where $\varphi_+ \gg \varphi_-$, while $\delta b \simeq 0$ because it is very massive, 
$m_b \gg H$. Thus, $\delta a_- \simeq -(\varphi_-/\varphi_+)\delta a_+$. Therefore, we have
\begin{equation}
\delta\theta_\pm = \frac{\delta a_\pm}{\sqrt{2}\varphi_\pm} 
\simeq \pm \frac{H}{2\sqrt{2}\pi \varphi_+}.
\end{equation}
The crucial point is that the amplitude of the fluctuation is determined solely by the larger field
value $\varphi_+$. Also notice that the fluctuations in the radial directions 
$\delta\varphi_+$ and $\delta\varphi_-$ are both suppressed due to large curvatures
in their potentials which stem from the first term in $V_{\rm SUSY}$ [Eq.(\ref{pot_susy})] and
the Hubble-induced mass terms [Eq.(\ref{pot-H})].

The axion isocurvature perturbation is given by\footnote{
The actual observable is 
${\cal S}_{\rm CDM} =(\Omega_a/\Omega_{\rm CDM}){\cal S}_a \propto \theta_+$,
where the axion density parameter is $\Omega_a\propto \theta_+^2$.}
\begin{equation}
{\cal S}_a \equiv \frac{\delta n_a}{n_a}-\frac{\delta n_\gamma}{n_\gamma}
=2\frac{\delta a}{a} \simeq \frac{H}{\sqrt{2}\pi \varphi_+ \theta_+},
\end{equation}
where $n_a$ and $n_\gamma$ denote the number densities of the axion 
and photon, respectively, and we use Eqs.(\ref{field}) and (\ref{delta_a}) in the last equality.
Therefore, the isocurvature fluctuation is written as
\begin{equation}
\Delta_{\cal S}^2(k)=A_{\rm iso}\left(\frac{k}{k_0}\right)^{n_{\rm iso}-1}, \
A_{\rm iso} \simeq \left.\frac{H^2}{2\pi^2\varphi_+^2\theta_+^2}\right|_{k=k_0}.
\end{equation}
Recent analyses of the admixture of adiabatic and isocurvature perturbations with 
independent spectral indices, $n_{\rm ad} \ne n_{\rm iso}$, reveal that the isocurvature
contribution can be as large as the adiabatic mode at the pivot scale $k_0$ and the blue spectral
index of the isocurvature fluctuation is favored such as $n_{\rm iso}\sim 4$ \cite{blue3}.

\section{Dynamics of the fields and isocurvature fluctuations}
As shown in the previous section, the amplitude of the isocurvature fluctuation is solely determined
by the larger field value $\varphi_+$ with the constant Hubble parameter during inflation, 
$H\simeq {\rm const}$. We therefore need to investigate the dynamics of $\varphi_+$ only.
It is reasonable to consider that the fields slide only along the flat direction, so that 
$\varphi_- = 2F_a^2/\varphi_+$, thus the potential can be approximated as
\begin{equation}
V  \simeq  \frac{1}{2} c_+H^2 \varphi_+^2 + 2c_-H^2F_a^4\frac{1}{\varphi_+^2}
\simeq  \frac{1}{2} c_+H^2 \varphi_+^2,
\end{equation}
where the last equality holds when $\varphi_+$ has a large field value.\footnote{
For large $\varphi_+$, kinetic terms are reduced to the normal one as 
$\displaystyle{\frac{1}{2}\sum_{i=\pm}\partial_\mu \varphi_i \partial^\mu \varphi_i =
\frac{1}{2}\left( 1+\frac{4F_a^4}{\varphi_+^4} \right)\partial_\mu \varphi_+ \partial^\mu \varphi_+
\simeq \frac{1}{2}\partial_\mu \varphi_+ \partial^\mu \varphi_+}$.}
Now we must just follow the same argument discussed in Sec.II. 
Since the $\varphi_+$ field obeys the equation
\begin{equation}
\ddot{\varphi}_+ + 3H\dot{\varphi}_+ + c_+ H^2 \varphi_+ = 0,
\end{equation}
whose solution is given by the form $\varphi_+ \propto e^{-\lambda H t}$ with
\begin{equation}
\lambda = \frac{3}{2} - \frac{3}{2}\sqrt{1-\frac{4}{9}c_+},
\end{equation}
the isocurvature fluctuation is obtained as
\begin{equation}
\Delta_{\rm iso}^2 \propto \left(\frac{\delta a}{\varphi_+}\right)^2 \sim \left(\frac{H}{\varphi_+}\right)^2
\propto e^{2\lambda H t},
\end{equation}
so that its spectral index becomes
\begin{equation}
n_{\rm iso} -1 \equiv \frac{d\ln\Delta_{\rm iso}^2}{d\ln k} = 2\lambda = 3 - 3\sqrt{1-\frac{4}{9}c_+}.
\end{equation}
Therefore, we obtain the blue spectrum with $1<n_{\rm iso} \le 4$ for $0 < c_+ \le 9/4$.
The prominent feature of this model is that $\varphi_+$ eventually settles down to the minimum
of the potential,
\begin{equation}
\label{phi_min}
\varphi_+^{\rm min} \simeq \sqrt{2} \left(\frac{c_-}{c_+}\right)^{1/4}\!\!F_a, \quad
\varphi_-^{\rm min} \simeq \sqrt{2} \left(\frac{c_+}{c_-}\right)^{1/4}\!\!F_a,
\end{equation}
and hence we have scale-invariant spectrum afterwards, smoothly connected from the blue 
spectrum at large scales. The e-folds during the field evolving from
$\simeq M_P$ to $\simeq F_a$ are estimated as
\begin{equation}
N_{\rm blue} \simeq \frac{1}{\lambda}\ln \left(\frac{M_P}{F_a}\right),
\end{equation}
which gives $N_{\rm blue} \simeq 10$ for $F_a=10^{12}$ GeV and $\lambda=3/2$ 
$(c_+=9/4)$, for example.

In order to confirm what we have obtained above, we solve numerically the equations for 
$\Phi_+$ and $\Phi_-$ with the potential (\ref{pot}). For the sake of numerical calculations, 
we decompose the field into real and imaginary parts as 
$\Phi_\pm = (\phi_\pm^{R} + i \phi_\pm^{I})/\sqrt{2}$, which leads to the following equations:
\begin{eqnarray}
&& \hspace*{-13mm} \ddot{\phi}_+^{R}+3H\dot{\phi}_+^{R}+c_+H^2\phi_+^{R} \nonumber \\
&&+\frac{h^2}{2}\left[ \left\{\phi_+^{R}\phi_-^{R} -\phi_+^{I}\phi_-^{I}-2F_a^2\right\}\phi_-^{R}
\right. \nonumber \\
&&\hspace{13mm}\left. + \left(\phi_+^{R}\phi_-^{I}+\phi_+^{I}\phi_-^{R}\right)\phi_-^{I}\right] =0,\\
&& \hspace*{-13mm} \ddot{\phi}_+^{I}+3H\dot{\phi}_+^{I}+c_+H^2\phi_+^{I} \nonumber \\
&&+\frac{h^2}{2}\left[ -\left\{\phi_+^{R}\phi_-^{R} -\phi_+^{I}\phi_-^{I}-2F_a^2\right\}\phi_-^{I}
\right. \nonumber \\
&&\hspace{13mm}\left. + \left(\phi_+^{R}\phi_-^{I}+\phi_+^{I}\phi_-^{R}\right)\phi_-^{R}\right] =0,\\
&& \hspace*{-13mm} \ddot{\phi}_-^{R}+3H\dot{\phi}_-^{R}+c_-H^2\phi_-^{R} \nonumber \\
&&+\frac{h^2}{2}\left[ \left\{\phi_+^{R}\phi_-^{R} -\phi_+^{I}\phi_-^{I}-2F_a^2\right\}\phi_+^{R}
\right. \nonumber \\
&&\hspace{13mm}\left. + \left(\phi_+^{R}\phi_-^{I}+\phi_+^{I}\phi_-^{R}\right)\phi_+^{I}\right] =0,\\
&& \hspace*{-13mm} \ddot{\phi}_-^{I}+3H\dot{\phi}_-^{I}+c_-H^2\phi_-^{I} \nonumber \\
&&+\frac{h^2}{2}\left[ -\left\{\phi_+^{R}\phi_-^{R} -\phi_+^{I}\phi_-^{I}-2F_a^2\right\}\phi_+^{I}
\right. \nonumber \\
&&\hspace{13mm}\left. + \left(\phi_+^{R}\phi_-^{I}+\phi_+^{I}\phi_-^{R}\right)\phi_+^{R}\right] =0.
\end{eqnarray}
%
\begin{figure}[ht]
\includegraphics[width=85mm]{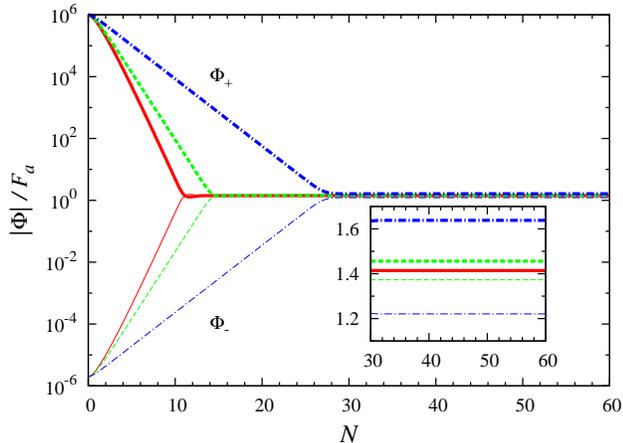}
\caption{Evolution of the fields $\Phi_+$ (upper thick lines) and $\Phi_-$ (lower thin lines) for
$c_-=9/4$ and $c_+=9/4$ ($n_{\rm iso}=4$, red solid), 2 ($n_{\rm iso}=3$, green dashed), 
and $5/4$ ($n_{\rm iso}=2$, blue dotted-dashed). The inset shows the minima where the fields 
settle down.}
\label{fig_evo}
\end{figure}

Some of the examples are shown in Fig.~\ref{fig_evo}. Here, we set the initial condition as
$|\Phi_+(0)|=M_P$. The initial values for $\Phi_-$ and the phases $\theta_\pm$ are taken so 
as to stay along the flat direction (\ref{FD}). We also take $h=1$ and $c_-=9/4$. 
In the figure, we show the results for $c_+=9/4$, $2$, and $5/4$, which correspond to 
$n_{\rm iso} = 4$, 3, and 2, respectively. One can see that the amplitude of the 
field $\varphi_+$ decreases exponentially, and eventually stays at the constant value, 
which coincides to Eq.(\ref{phi_min}). This is very attractive, since there is no blow-up 
of the spectrum at smaller scales, while having extremely blue tilt even as large as 
$n_{\rm iso}=4$ at large scales over a few orders of magnitude. Notice that the 
result remain unchanged even if we vary the Hubble parameter provided that  
$H\ll F_a$; here, we take a particular value as $H/F_a=10^{-2}$.

Finally we comment on the initial amplitude of the fields. The $\varphi_+$ should be at large 
field values in the beginning. One of the simple mechanism to realize this situation is to consider 
pre-inflation, where the pre-inflaton and the $\Phi_+$ have nonrenomalizable coupling in 
the K\"ahler potential so as to give rise to a negative Hubble-induced mass term 
during pre-inflation. In this case, the initial condition will be $\varphi_+ \simeq M_P$.

\section{Conclusions}
We have proposed the concrete model for generating isocurvature fluctuations with 
extremely blue spectrum for some range of the scale. It is based on the axion
model in supersymmetry. The supergravity effects raise the Hubble-induced mass
terms in the potential of the $\varphi_\pm$ fields.
These Hubble-induced mass terms play two roles. One is that they suppress the
fluctuations in the radial directions, $\delta\varphi_\pm$. The other is to make the fields
evolve during inflation. In particular, the field value of $\varphi_+$ determines
the amplitude of the axion isocurvature perturbation: the blue tilt is due to the dynamics
of $\varphi_+$ moving from the large initial value ($\sim M_P$) down to the PQ symmetry
breaking scale $F_a$ during inflation.
Depending on the coefficient of the Hubble-induced mass term, $c_+$, we can obtain
$1< n_{\rm iso}\le4$. The prominent feature of this model is that the blue spectrum is 
realized only while $\varphi_+$ is evolving and after it settles down into the potential 
minimum the spectrum becomes scale invarinat.

The actual scale $L$ where the spectrum is blue is determined by e-folds $N'$ after 
$\varphi_+$ settles down to $F_a$. For example, $n_{\rm iso} > 1 $ at 
$L \gtrsim 1$~Mpc for $N' \simeq 47$ assuming that the present Hubble radius 
corresponds to $N \simeq 55$. Observations of large-scale structures, or even 
PLANCK, could see the existence of the isocurvature fluctuations with a huge 
blue tilt, which may approve our model in the near future.

\section*{Acknowledgments}
The authors are grateful to Toyokazu Sekiguchi for discussion. 
This work is supported by Grant-in-Aid for Scientific research from the Ministry of 
Education, Science, Sports, and Culture (MEXT), Japan, under Contract No. 14102004 (M.K.),
and also by World Premier International Research Center Initiative, MEXT, Japan.



\end{document}